\documentclass[referee]{cjaa}
\usepackage{graphicx}
\input{epsf.sty}
\input{psfig.sty}                       

\begin{document}

\title{A Super-high Angular Resolution Principle for Coded-mask X-ray Imaging Beyond the Diffraction Limit of Single Pinhole}

   \volnopage{Vol.0 (200x) No.0, 000--000}      
   \setcounter{page}{1}          

\author{Chen Zhang\inst{1} \and Shuang Nan Zhang \inst{2,3}}
\institute{Department of Engineering Physics \& Center for Astrophysics, Tsinghua
University, Beijing 100084, P. R.
China;\\
\email{zhangsn@tsinghua.edu.cn} \and Key Laboratory of Particle Astrophysics, Institute
of High Energy Physics, the Chinese Academy of Sciences, Beijing 100049, P.
R. China;\\
\and Department of Physics \& Center for Astrophysics, Tsinghua University, Beijing
100084, P. R.
China;\\
}

   \date{Received~~2007 month day; accepted~~2007~~month day}

\abstract{High angular resolution X-ray imaging is always demanded by astrophysics and
solar physics, which can be realized by coded-mask imaging with very long mask-detector
distance in principle.  Previously the diffraction-interference effect has been thought
to degrade coded-mask imaging performance dramatically at low energy end with very long
mask-detector distance. In this work the diffraction-interference effect is described
with numerical calculations, and the diffraction-interference cross correlation
reconstruction method (DICC) is developed in order to overcome the imaging performance
degradation. Based on the DICC, a super-high angular resolution principle (SHARP) for
coded-mask X-ray imaging is proposed. The feasibility of coded mask imaging beyond the
diffraction limit of single pinhole is demonstrated with simulations. With the
specification that the mask element size of $\mathrm{50 \times 50  \ \mu m^2}$ and the
mask-detector distance of 50 m, the achieved angular resolution is 0.32 arcsec above
about 10 keV, and 0.36 arcsec at 1.24 keV ($\lambda =1$ nm) where diffraction can not
be neglected. The on-axis source location accuracy is better than 0.02 arcsec.
Potential applications for solar observations and wide-field X-ray monitors are also
shortly discussed.
}

   \authorrunning{C. Zhang \& S. N. Zhang }            
   \titlerunning{SHARP - A super-high angular resolution principle}  

     \maketitle

\section{Introduction}
\label{sect:intro}
High angular resolution X-ray imaging is always demanded by astrophysics and solar
physics, for example to study black holes near event horizon, relativistic jets of
super massive black holes, as well as solar flares and coronal activities. So far the
best imaging technology for X-ray observation is realized by grazing incidence
reflection (Aschenbach~\cite{Aschenbach1985}), which provides a very good angular
resolution, for example down to 0.5 arcsec as in the case of the Chandra X-ray
Observatory (Weisskopf et al.~\cite{Weisskopf2000}). The diffraction limit of Chandra
is about 8 miliarcsec at 6 keV; however slope errors and other surface irregularities
that affect the angle of reflection prevent the angular resolution better than 0.5
arcsec. Further more, the grazing incidence reflection is limited by the working energy
band, which hardly exceeds tens of keV. A possible way to improve the angular
resolution beyond Chandra's requires a different technology, for example the
diffractive-refractive X-ray optics proposed by Gorenstein\ (\cite{Gorenstein2007}) or
the Fourier-Transform imaging by Prince et al.\ (\cite{Prince1988}). As an alternative,
a super-high angular resolution principle (SHARP) for a coded-mask X-ray imaging
telescope concept is proposed here.

The coded mask imaging technology, which has been reviewed by Zand\
(\cite{Zand19921996}), is widely applied for X-ray observations, for example the
INTEGRAL mission (Winkler et al.~\cite{Winkler2003}) of ESA and the SWIFT mission
(Gehrels~\cite{Gehrels2004}) of NASA. Both missions have angular resolution in tens of
arcmin, since the distances between the masks and the detectors are limited by the
dimensions of the satellites. In principle super-high angular resolution better than
Chandra can be achieved by a coded-mask telescope with sub-millimeter size of mask
pinholes and  long mask-detector distances, which can be  tens, even hundreds of meters
with the formation-flying technology (for example proposed for XEUS
(ESA~\cite{XEUS2001}) of ESA) or the mast technology (for example applied in Polar
mission of NASA). The X-ray diffraction effect in a coded-mask system has been commonly
thought to be negligible due to high photon energies in hard X-ray and gamma-ray range
as well as relatively small mask-detector distance. Whereas for coded-mask telescope
with a super high angular resolution in the milli-arc-second range, the X-ray
diffraction effect could not be neglected, especially at low energy end (in the range
of several keV), which might degrade the imaging performance remarkably, as pointed out
by Prince et al.\ (\cite{Prince1988})  and Skinner\ (\cite{Skinner2004}). In this work,
we study the diffraction effect in SHARP with numerical computations, and propose a
diffraction-interference cross correlation reconstruction method and demonstrate the
feasibility of coded mask imaging beyond the diffraction limit of single pinhole.  A
potential application for solar observations is also shortly discussed.

\section{The coded-mask principle and diffraction-interference reconstruction}
\subsection{The coded-mask principle and SHARP}
\begin{figure}
   \centering
   \includegraphics[width=0.5\textwidth]{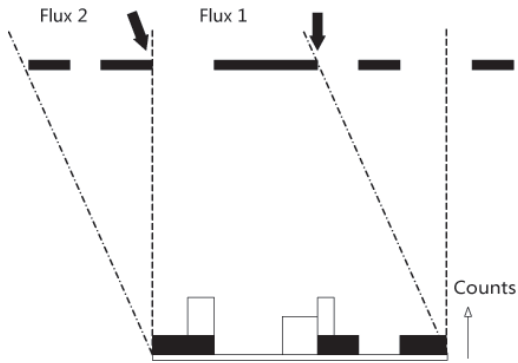}
   \caption{The basic concept of coded-mask imaging, in the limit of geometrical optics, i.e.,
 the diffraction effect is negligible.
 Two point sources illuminate a
position-sensitive detector plane through a mask made of many square pinholes. The
detector plane thus records two projections of the mask pattern. The shift and the
strength of projections encode the position and the flux of the sources separately.
The PSF width is defined as $\Delta_i=\frac{d_m}{D}$. }
   \label{fig_principle}
   \end{figure}

In the limit of geometrical optics, i.e., the diffraction effect is not considered, the
basic concept of coded-mask imaging is shown in Fig. \ref{fig_principle}. The
coded-mask camera has a mask on top of a position-sensitive detector with a
mask-detector distance $D$. Two point sources project the mask pattern onto the
detector plane. The shift and the strength of projections encode the position and the
flux of sources separately.  The detection of the X-ray flux can be described with Equ.
\ref{equ_x_detect} (Fenimore et al.~\cite{E.E.Fenimore1978}),
\begin{equation}
\label{equ_x_detect} P=O*M+N_{\mathrm{noise}},
\end{equation}
where $O$ is the X-ray flux spatial distribution, $M$ is the encoding pattern of mask
and $*$ means cross correlation. For the commonly applied cross correlation method, a
matrix $G$ is used for reconstruction, as shown in Equ.  \ref{equ_x_recon},
\begin{equation}
\label{equ_x_recon} O'=P*G=O*(M*G)+N_\mathrm{{noise}}*G,
\end{equation}
where $O'$ is the estimation of X-ray flux spatial distribution, if $M*G$ is a $\delta$
function. The encoding pattern $M$ is often chosen so that its auto-correlation
function is a $\delta$ function. Therefore $G=M$ is commonly applied. The width of the
Point Spread Function (PSF) obtained with Equ.  \ref{equ_x_recon} is defined as
$\Delta_i=\frac{d_m}{D}$ ($d_m$ is the size of the mask element).  The cyclic optimal
coded-mask configuration (Zand~\cite{Zand19921996}) is also employed here. The coded
mask or the mask pattern consisted of four basic patterns is about
 four times the size of the detector, and one basic pattern cyclically extends about twice along the width and
 length directions of the mask. In the next sections, simulations are done with the following assumptions: the basic pattern is totally random; the detector pixel size is equal to the mask element size, both of which are in square shape; the mask and the detector are assumed to be idealized, i.e., the mask has no thickness, the opaque mask elements block the X-ray completely and one incident event can only be recorded in one detector pixel.

In this work, $d_m=50 \ \mathrm{\mu m}$ and $D=50 \ \mathrm{m}$ are chosen to be the
basic parameters for the simulations of SHARP, which means  $\Delta_i=0.2 \
\mathrm{arcsec}$ in sub-arcsec range. Since the simulations employ two dimensional FFT
(please refer to Sec. \ref{sec_diffraction}) which requires tremendous computing
resources, the basic pattern simulated contains only $200 \times 200$ elements, i.e.,
the reconstructed image contains the same pixels and each pixel is $0.2 \times 0.2 \
\mathrm{arcsec^2}$. The fully coded Field-Of-View (FOV) is $40 \times 40 \
\mathrm{arcsec^2}$.

\subsection{The diffraction-interference cross correlation method }
\label{sec_diffraction} 
\begin{figure}
   \centering
   \includegraphics[width=0.8\textwidth]{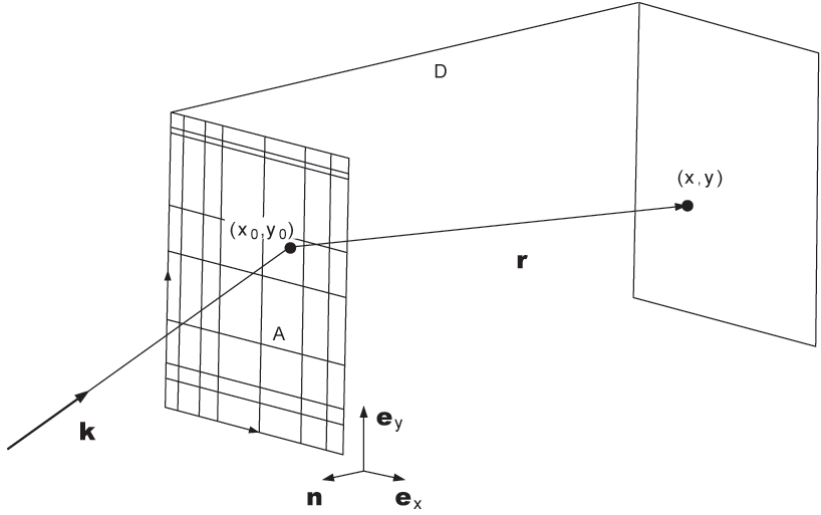}
   \caption{Monochromatic X-ray with wave vector $\mathbf{k}$ illuminating onto a mask in the $\mathrm{x_0}$-$\mathrm{y_0}$ plane produces a diffraction pattern in the x-y plane  (Lindsey~\cite{Lindsey1978}). }
   \label{fig_fel_principle}
   \end{figure}

As shown in Fig. \ref{fig_fel_principle}, $A$ is the mask
modulation function, which is 1 or 0 with idealized mask. $\mathbf{r}$ is a vector from
point $(x_0,y_0)$ on the mask to the $(x,y)$ on the detector. $\mathbf{k}$ is the wave
vector. Since there are many pinholes in the mask, both diffraction and interference
will take place on the detector plane. The diffraction-interference is described with
Fresnel-Huygens principle (Lindsey~\cite{Lindsey1978}) as,
\begin{equation}
\label{equ_fenir} E_R(x,y)=C\int_\infty ^ \infty  \int_\infty ^ \infty
E_i(x_0,y_0)A(x_0,y_0)\frac{e^{ikr}}{r} dx_0 dy_0,
\end{equation}
where $E_i$ is the amplitude distribution of incident photons, which is described with
the plane wave in our case as $E_i=A_i*e^{i\mathbf{k \cdot r}}$. The flux distribution
is proportional to $E_i^2$.  $C=-i/\lambda$ is constant. In our case the dimension of
the mask is far less than the mask-detector distance $D$, we have  $r \simeq
D+\frac{x^2+x_0^2+y^2+y_0^2-2xx_0-2yy_0}{2D}$. Therefore  Equ. \ref{equ_fenir} becomes
\begin{equation}
\label{equ_fenir_fourier} E_R(x,y)= C_2\int_\infty ^ \infty  \int_\infty ^ \infty M
e^{-2i\pi (f_x x_0+f_y y_0)} dx_0 dy_0,
\end{equation}
where ~$C_2=\frac{-i}{\lambda D} e^{ikD} e^{\frac{ik}{2D}(x^2+y^2)}$, $f_x=x/\lambda
D$, $f_y=y/\lambda D$, $M=E_i(x_0,y_0)A(x_0,y_0)e^{\frac{ik}{2D}(x_0^2+y_0^2)}$. Equ.
\ref{equ_fenir_fourier} is the Fourier transform approximation of Fresnel-Huygens
principle, which can be simulated with an FFT algorithm. The conditions for the
validity of this approximation are discussed by Lindsey\ (\cite{Lindsey1978}), and are
satisfied in our simulations.

\begin{figure}
   \centering
   \includegraphics[width=0.6\textwidth]{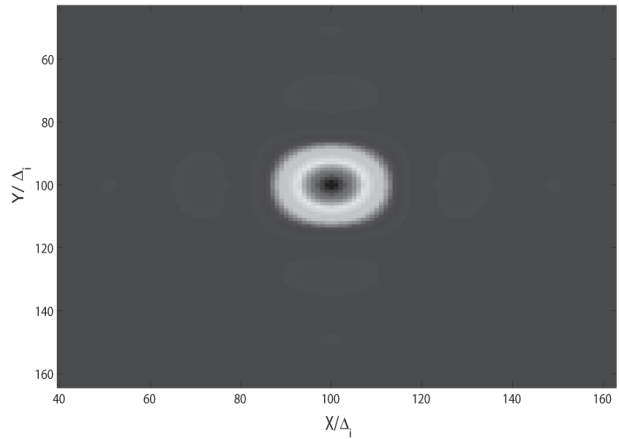}
   \caption{The simulated flux distribution on the detector plane for 1.24 keV photons ($\lambda=1$ nm)
   with a $50 \ \mathrm{\mu m}$ square pinhole in front of it (50 m away).
   The flux spreads much larger than $50 \ \mathrm{\mu m}$ caused by
   diffraction as one pixel is $50 \times 50 \ \mathrm{\mu m^2}$.
   Therefore this diffraction effect must be included when reconstructing the image of
   the incoming X-ray with cross-correlation effect.}
   \label{fig_diff}
   \end{figure}

 The X-ray diffraction can not be neglected at the low energy end as in Fig. \ref{fig_diff}, which shows the simulated flux distribution on the detector plane with a $50 \ \mathrm{\mu m}$ square pinhole in front of it (50 m away) for 1.24 keV photons ($\lambda=1$ nm). The flux spreads much larger than $50 \ \mathrm{\mu m}$ (each pixel in Fig. \ref{fig_diff} is $50 \times 50 \ \mathrm{\mu m^2}$) due to the diffraction effect.

\begin{figure}
   \centering
   \includegraphics[width=\textwidth]{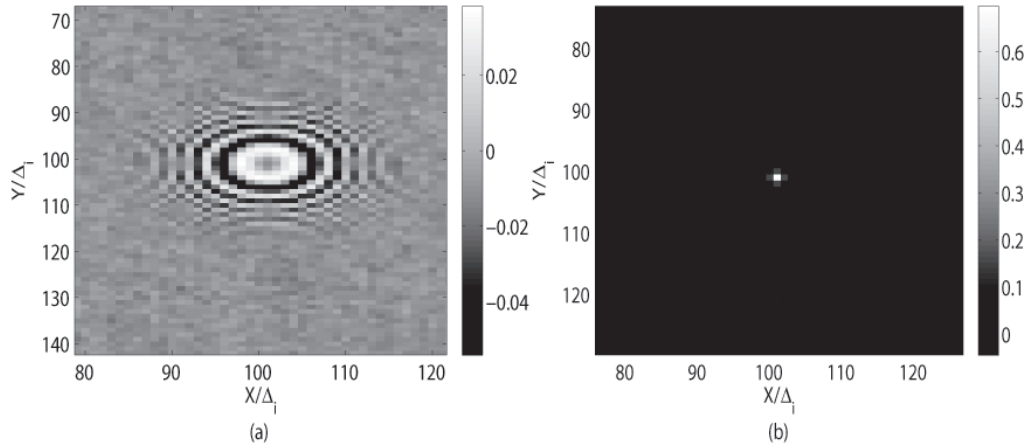}
   \caption{The reconstructed images of a monochromatic ($\lambda=1$ nm)
   on-axis point source: (a) the cross-correlation with the $G_O$,
   the diffraction degrades the reconstruction dramatically, because the diffraction pattern shown in Fig. 3 has not
   been taken into consideration;
   (b) with DICC, the point source can be clearly identified although slight Fresnel diffraction stripes still exist. The
   DICC can thus largely overcome the degradation of the imaging performance due to the diffraction of each pinhole.}
   \label{fig_recon_opt_diff}
   \end{figure}

 If the reconstruction matrix $G$ in the cross correlation method is still chosen to be equal to $M$ (so-called geometrical reconstruction matrix, which is indicated as $G_O$) at the low energy end, the diffraction effect would degrade the reconstructed image remarkably. Fig. \ref{fig_recon_opt_diff}a shows the reconstructed image of a monochromatic on-axis point source ($\lambda = 1 \ nm$) obtained by geometrical matrix. The reconstructed image is normalized so that the total integral flux is one photon, and all the reconstructed images in this work are normalized in the same way if not specified. The point source is hardly identified with the Fresnel diffraction stripes clearly observed. Since the X-ray probability wave through different mask open elements interferes with each other, the mask spatial information should be encoded into the diffraction-interference pattern on the detector. We then choose the reconstruction matrix $G$ at certain energy as the normal incident diffraction-interference pattern (so-called diffraction-interference matrix, which is indicated as $G_{DI}$) on the detector plane. Fig. \ref{fig_recon_opt_diff}b shows the reconstructed image obtained with $G_{DI}$ of the same source in Fig. \ref{fig_recon_opt_diff}a, and the point source can be clearly identified although slight Fresnel stripes still exist. This new reconstruction method is called diffraction-interference cross correlation method (DICC), which can be described with Equ.  \ref{equ_x_dicc},
\begin{equation}
\label{equ_x_dicc} O'=P*G_{DI}=O*(M*G_{DI})+N_\mathrm{{noise}}*G_{DI}.
\end{equation}
$G_{DI}$ can be obtained by numerical simulation as shown in Equ.
\ref{equ_fenir_fourier} or from actual measurements. The angular resolution of DICC is
beyond the single pinhole (the single mask open element) diffraction limit as indicated
in Fig. \ref{fig_recon_opt_diff}.

\section{The simulation results}

We have simulated the reconstructed images at several energies from 1.24 keV
($\lambda=1$ nm) to 6.2 keV ($\lambda=0.2$ nm), where DICC is valid. To illustrate the
main properties of this method, we mainly discuss the imaging performance of DICC
around 1 nm, which is also compared with that in the limit of geometrical optics.

\subsection{Angular resolution}
\begin{figure}
   \centering
   \includegraphics[width=\textwidth]{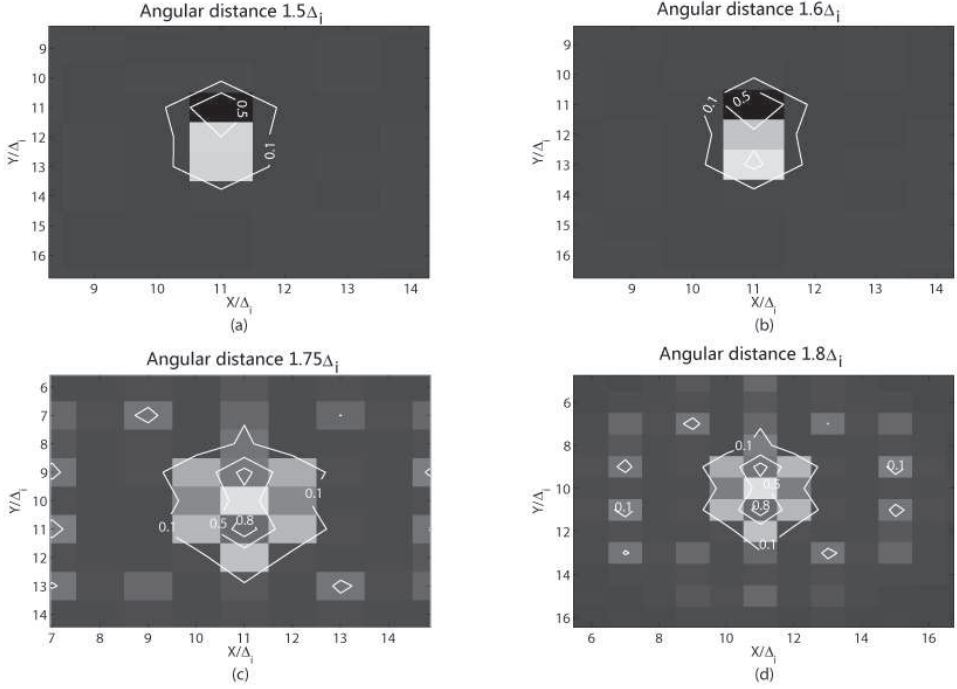}
   \caption{The angular resolution of the simulated system: (a) and (b) are in the limit of geometrical optics;  (c) and (d) are obtained with DICC for $\lambda=1$ nm sources. In (a) and (c) two sources can not be separated. However in (b) and (d) two sources can be separated. The angular resolution is 0.32 arcsec in the limit of geometrical optics and degraded to 0.36 arcsec for  $\lambda=1$ nm sources.}
   \label{fig_an_res}
   \end{figure}

We calculate the reconstructed image of two point sources with the same flux (Poisson
fluctuations are not considered here). The angular resolution is defined as the minimum
angular distance between these two sources, which can be separated by the 50\%  contour
line of the maximum flux on the reconstructed image. Fig. \ref{fig_an_res}a shows the
reconstructed images in the limit of geometrical optics for two point sources with
angular distance $1.5 \Delta _i=0.3 \ \mathrm{arcsec}$; the two sources can not be
separated. However in Fig \ref{fig_an_res}b two sources with angular distance $1.6
\Delta _i=0.32 \ \mathrm{arcsec}$ can just be separated. Therefore the angular
resolution is 0.32 arcsec in the limit of geometrical optics, i.e., at high energy end
(above about 10 keV) with diffraction negligible. With the same procedure, the angular
resolution of DICC for $\lambda=1$ nm photons is obtained as 0.36 arcsec, i.e., the two
sources can not be separated in Fig. \ref{fig_an_res}c, but can be separated in Fig.
\ref{fig_an_res}d. This means the diffraction effect degrades the angular resolution by
only about 12.5\%, a significant improved compared to that in Fig. 3. We also
calculated the reconstructed images with DICC for $\lambda=1$ nm point sources at
different locations inside the FOV, but no observable reconstruction degradation is
found compared with the on-axis source case.

\subsection{Source location accuracy}

\begin{figure}
   \centering
   \includegraphics[width=\textwidth]{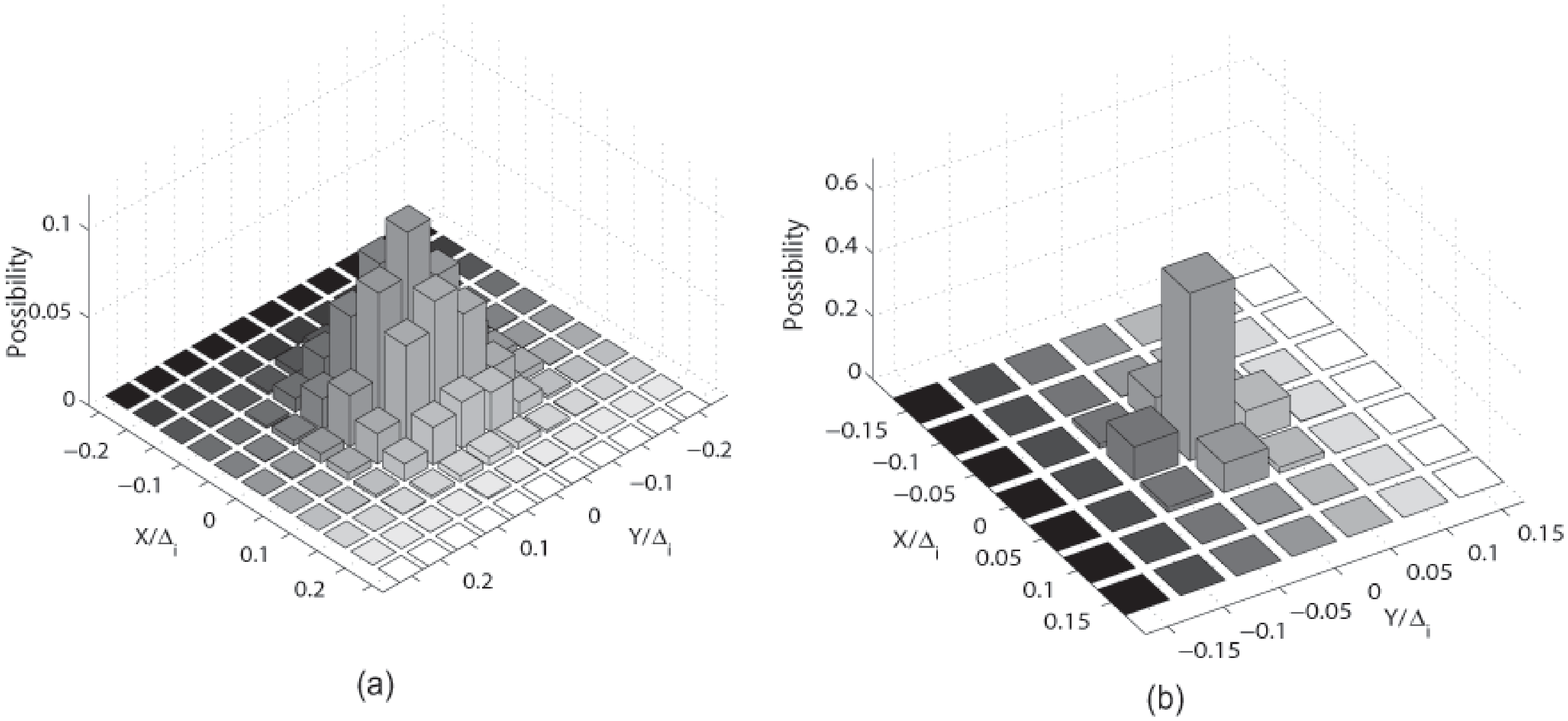}
   \caption{The calculated source location distribution of the reconstructed images for an  on-axis point source, which is simulated 10000 times with 1000 photons recorded: (a) with DICC for $\lambda=1$ nm photons; (b) in the limit of geometrical optics. The source location accuracy is better than $0.1 \Delta _i=0.02 \ \mathrm{arcsec}$.}
   \label{fig_local}
   \end{figure}

The reconstructed images for an on-axis point source is simulated 10000 times with 1000
photons recorded (Poisson fluctuations are considered). The source location of each
simulation is calculated by fitting the reconstructed image to the PSF. The calculated
source location distribution with DICC for $\lambda=1$ nm is shown in Fig.
\ref{fig_local}a, which is concentrated within $0.1 \Delta _i=0.02 \ \mathrm{arcsec}$.
The same simulation is also done in the limit of geometrical optics, which is shown in
Fig. \ref{fig_local}b. Therefore the source location accuracy is better than $0.1
\Delta _i=0.02 \ \mathrm{arcsec}$ for 1000 detected photons.

\subsection{System constraints}

\begin{figure}
   \centering
   \includegraphics[width=\textwidth]{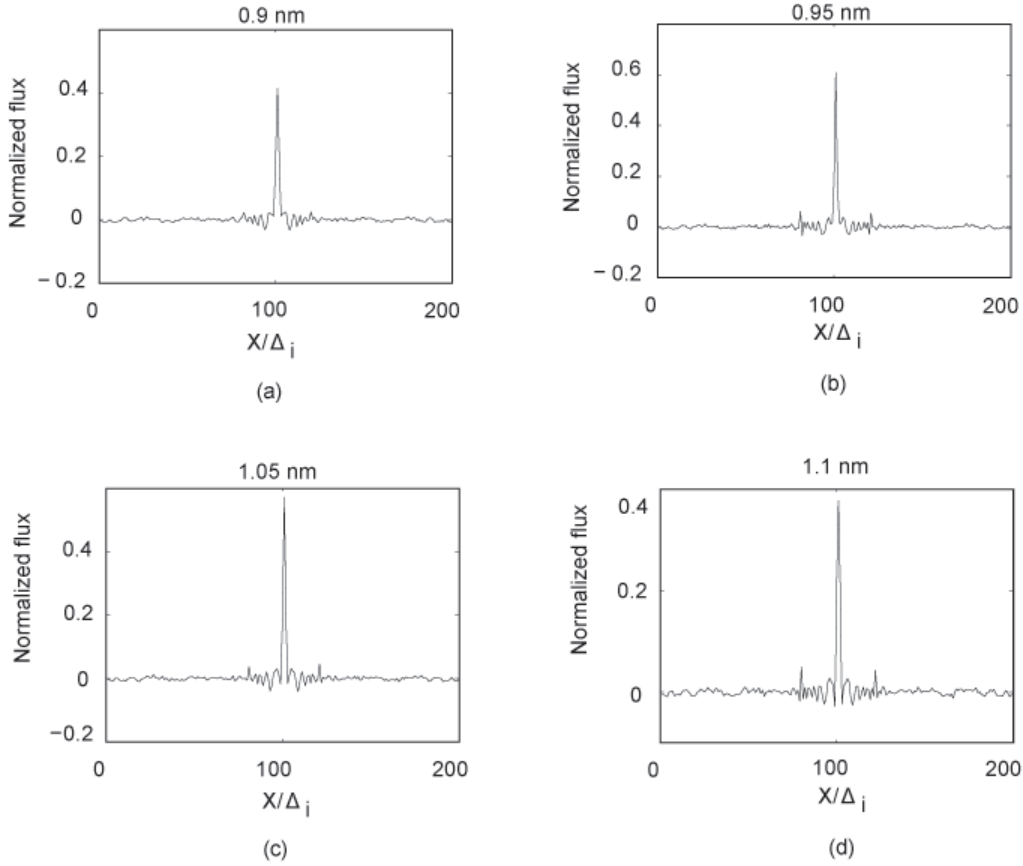}
   \caption{The reconstructed images of (a) 0.9 nm (b) 0.95 nm (c) 1.05 nm and (d) 1.1 nm photons, which all apply the same diffraction-interference matrix $G_1$ (the diffraction-interference matrix of 1 nm photons). }
   \label{fig_cons}
   \end{figure}

Since the diffraction pattern on the detector plane is energy dependent, the
diffraction matrix $G$ is also energy dependent. In practice the detector can not
identify two photons with an energy difference within the detector's spectroscopy
resolution, i.e., in practice photons within a small energy band must share a common
diffraction matrix. Fig. \ref{fig_cons} shows the reconstructed images of several
on-axis point sources with wavelengths around 1 nm, which all apply the diffraction
matrix of 1 nm photons (marked as $G_1$). The point source can be clearly identified.
Obviously the imaging qualities are degraded by applying the $G_1$ instead of their own
diffraction matrixes. Tab. \ref{tab_cons_anr} gives the angular resolution at several
wavelengths around 1 nm by applying $G_1$ as reconstruction matrix instead of their
own. For the 1.05 nm on-axis source, the angular resolution is degraded to $2.05\Delta
_i=0.41 \ \mathrm{arcsec}$ by applying $G_1$.

 \begin{table}[]
  \caption[]{The Degradation of the Angular Resolution with Shared Reconstruction Matrix}
  \label{tab_cons_anr}
  \begin{center}\begin{tabular}{clclclclclclclcl}
  \hline\noalign{\smallskip}
Wavelength (nm)  & 0.95 &0.97&0.99&1&1.01&1.03&1.05                    \\
  \hline\noalign{\smallskip}
Angular resolution (arcsec)  & 0.39     & 0.38&0.37&0.36&0.37&0.39&0.41        \\
  \noalign{\smallskip}\hline
  \end{tabular}\end{center}
\end{table}

Based on the detector energy resolution, the system constraint is considered. Assuming
the detector pixel size $d_d$, the difference $\delta d$ of the main diffraction stripe
diameters  of photons with wavelength difference $\delta \lambda$ is
\begin{equation}
\delta d=\frac{\delta \lambda*D}{d_m}.
\end{equation}

If the spectroscopy resolution of the detector is $\delta \lambda$, spatially the
detector does not need to distinguish $\delta d$ for photons with wavelength difference
$\delta \lambda$,
\begin{equation}
\delta d \leq f*d_d.
\end{equation}
Here $f\ge 1$ is a constant and depends upon the mask open fraction. For a 50\% open
mask, our simulations indicates that $1\le f\le 2$. For a mask fraction much less than
50\%, $f>2$ is possible. Then with known basic parameters, i.e., $d _m$ and $D$, the
requirement for detector spectroscopy resolution is
\begin{equation}
\label{equ_sec_x_sys_cons_en} \delta \lambda \leq f*\Delta _i*d_d.
\end{equation}
In practice $d _m$ and $D$ are chosen to match the detector performance, which means
\begin{equation}
\label{equ_sec_x_sys_cons_dm} d_m \geq \frac{\delta \lambda}{f *\Delta _i},
\end{equation}
or
\begin{equation}
\label{equ_sec_x_sys_cons_D} D=\frac{d_m}{\Delta _i} \leq  \frac{d_d}{\Delta _i}\leq
\frac{\delta\lambda}{(\Delta _i)^2f}.
\end{equation}

Therefore the larger the $f$ is, the more compact the system can be. Here $f=1$ is
chosen for convenience, with which $\delta \lambda=0.05$ nm, $d_m=50\ \mathrm{\mu m}$
and $D=50$ m satisfies the system constraint. The spectroscopy resolution 0.05 nm
requires about 120 eV energy resolution @$\lambda=1$ nm, which can be provided by
modern silicon imaging detectors like CCD or DEPFET (Str\"uder~\cite{Struder2000}).

\section{Discussion and Conclusion}
By applying DICC, the system angular resolution is no more limited by single pinhole
diffraction limit at low energy end, but limited by detector spectroscopy performance.
Tab. \ref{tab_com} shows a brief comparison of SHARP with several X-ray astronomy
missions. The advantage of SHARP is quite obvious: excellent angular resolution and
easily fabricated mask which does not require high accuracy mechanical fabrication of
X-ray grazing lens. However the detector effective area of a coded-mask telescope is
smaller than the its actual area due to the encoding pattern. Further more each
detector pixel receives background photons from all over the FOV. Therefore the
sensitivity is limited by its large background and small effective area compared with
that in the grazing incidence reflection case.

Current detector technology can provide even better energy resolution than 0.05 nm @1
nm, which means the angular resolution discussed above can be improved further by
increasing the mask-detector distance $D$ as indicated in Equ.
\ref{equ_sec_x_sys_cons_D} or by decreasing the mask element size as indicated in Equ.
\ref{equ_sec_x_sys_cons_dm}. For example for a detector with 0.002 nm spectroscopy
resolution (5 eV @ 1 keV), $d_m=40\ \mathrm{\mu m}$ and $D=800$ m satisfies the system
constraint with $\Delta _i=0.01 \ \mathrm{arcsec}$.

A potential application of SHARP will be the solar observation at 1-100 keV energy band
with sub-arcsec angular resolution. The main objectives may be the coronal mass
ejection (CME) and solar flares, including  fine structures and evolutions of the solar
flares, nonlinear solar flare dynamics, solar particle acceleration mechanism et al.
Therefore applications of SHARP are foreseen to make significant progress on the study
of solar high energy explosive events and the space weather forecast model. Another
potential application is to make wide-field X-ray monitors with SHARP; its sub-arc
angular resolution may allow the counterparts of gamma-ray bursts, X-ray bursts, black
hole and neutron star transients to be identified without the requirement of subsequent
follow-up observations of focusing X-ray telescopes.


\begin{table}
 \caption{The Comparison Between SHARP and Several Missions }
 \begin{minipage}[t]{0.8\linewidth}
\label{tab_com}
  \begin{center}\begin{tabular}{p{1.5cm}p{1.5cm}p{1.5cm}p{1.5cm}p{1.5cm}p{1.5cm}}
  \hline\noalign{\smallskip}
 Mission &SHARP & Chandra  & Integral\footnote{IBIS onboard} &RHESSI  & Hinode (Solar B)\footnote{XRT onboard}\\
  \hline\noalign{\smallskip}
Imaging technology&Coded-mask&Focusing&Coded-mask&Rotation Modulation \& Fourier Transform&Focusing\\
Energy band & 1-100 keV&Up to 10 keV&15 keV-10 MeV&3-150 keV\footnote{The upper layer}&About to 6 keV\\
Energy  resolution& 133 eV@5.9 keV\footnote{Take DEPFET as focal plane detector (Treis et al.~\cite{Treis2006})} &ACIS: about 148 eV@5.9 keV\footnote{CCD S3 on board}&9 \% @ 100 keV&$<1$ keV\footnote{Below 100 keV}&$<280$ eV @5.9 keV\footnote{Calculated based on the ENC $<30$ el.} \\
Angular resolution &0.32-0.36 arcsec&0.5 arcsec\footnote{On-axis point source}&12 arcmin&2.26 arcsec\footnote{The collimator \# 1}&2 arcsec @ 0.523keV\footnote{68\% source flux within 2 arcsec}\\
  \noalign{\smallskip}\hline
  \end{tabular}\end{center}

\end{minipage}

\end{table}

\begin{acknowledgements}
We thank C. Fang, W. Q. Gan, J. Y. Hu, Z. G. Dai, X. D. Li, Y. F. Huang and X. Y. Wang
for many interesting discussions and suggestions on SHARP and its potential
applications. We are grateful to the anonymous referee for providing valuable comments
and suggestions promptly. SNZ acknowledges partial funding support by the Ministry of
Education of China, Directional Research Project of the Chinese Academy of Sciences
under project No. KJCX2-YW-T03 and by the National Natural Science Foundation of China
under grant Nos. 10521001, 10733010, 10725313 and 10327301.

\end{acknowledgements}

\clearpage

\end{document}